\documentclass[sn-mathphys]{sn-jnl}% Math and Physical Sciences Reference Style
\jyear{2021}%
\usepackage{natbib}
\usepackage{booktabs}
\usepackage{makecell}
\usepackage{bm} 
\usepackage{hhline}
\usepackage{colortbl}
\usepackage{multicol}

\theoremstyle{thmstyleone}%
\theoremstyle{thmstyletwo}%

\theoremstyle{thmstylethree}%
\usepackage{subfig}
\begin{document}

\title[Improving repeatability with Monte Carlo dropout]{Improving the repeatability of deep learning models with Monte Carlo dropout}


\author[1,2]{\fnm{Andreanne} \sur{Lemay}}\email{andreanne.lemay@polymtl.ca}
\author[1,3]{\fnm{Katharina} \sur{Hoebel}}\email{khoebel@mit.edu}
\author[1,4]{\fnm{Christopher P.} \sur{Bridge}}\email{cbridge@partners.org}
\author[5]{\fnm{Brian} \sur{Befano}}\email{befanob@uw.edu}
\author[6]{\fnm{Silvia} \sur{De Sanjosé}}\email{desanjose.silvia@gmail.com}
\author[6]{\fnm{Didem} \sur{Egemen}}\email{didem.egemen@nih.gov}
\author[6]{\fnm{Ana Cecilia} \sur{Rodriguez}}\email{rodriguezac2@gmail.com}
\author[6]{\fnm{Mark} \sur{Schiffman}}\email{schiffmm@exchange.nih.gov}
\author[7]{\fnm{John Peter} \sur{Campbell}}\email{campbelp@ohsu.edu}
\author*[1]{\fnm{Jayashree} \sur{Kalpathy-Cramer}}\email{kalpathy@nmr.mgh.harvard.edu}

\affil*[1]{\orgdiv{Martinos Center for Biomedical Imaging}, \orgaddress{\city{Boston}, \state{MA}, \country{USA}}}

\affil[2]{\orgdiv{NeuroPoly}, \orgname{Polytechnique Montreal}, \orgaddress{\city{Montreal}, \state{QC}, \country{Canada}}}

\affil[3]{\orgdiv{Massachusetts Institute of Technology}, \orgaddress{\city{Cambridge}, \state{MA}, \country{USA}}}

\affil[4]{\orgdiv{MGH \& BWH Center for Clinical Data Science}, \orgaddress{\city{Boston}, \state{MA}, \country{USA}}}

\affil[5] {\orgdiv{Department of Epidemiology}, \orgname{University of Washington School of Public Health}, \orgaddress{\city{Seattle}, \state{WA}, \country{USA}}}

\affil[6]{\orgdiv{Division of Cancer Epidemiology {\&} Genetics}, \orgname{National Cancer Institute}, \orgaddress{\city{Rockville}, \state{MD}, \country{USA}}}

\affil[7]{\orgname{Oregon Health and Science University}, \orgaddress{\city{Portland}, \state{OR}, \country{USA}}}

\abstract{The integration of artificial intelligence into clinical workflows requires reliable and robust models. Repeatability is a key attribute of model robustness. Repeatable models output predictions with low variation during independent tests carried out under similar conditions. During model development and evaluation, much attention is given to classification performance while model repeatability is rarely assessed, leading to the development of models that are unusable in clinical practice. In this work, we evaluate the repeatability of four model types (binary classification, multi-class classification, ordinal classification, and regression) on images that were acquired from the same patient during the same visit. We study the performance of binary, multi-class, ordinal, and regression models on four medical image classification tasks from public and private datasets: knee osteoarthritis, cervical cancer screening, breast density estimation, and retinopathy of prematurity. Repeatability is measured and compared on ResNet and DenseNet architectures. Moreover, we assess the impact of sampling Monte Carlo dropout predictions at test time on classification performance and repeatability. Leveraging Monte Carlo predictions significantly increased repeatability for all tasks on the binary, multi-class, and ordinal models leading to an average reduction of the 95\% limits of agreement by 16\% points and of the disagreement rate by 7\% points. The classification accuracy improved in most settings along with the repeatability. Our results suggest that beyond about 20 Monte Carlo iterations, there is no further gain in repeatability. In addition to the higher test-retest agreement, Monte Carlo predictions were better calibrated which leads to output probabilities reflecting more accurately the true likelihood of being correctly classified.}

\keywords{repeatability, Monte Carlo dropout, cervical screening, breast density, retinopathy of prematurity, knee osteoarthritis, medical classification, computer vision}

\maketitle

\section{Introduction}\label{sec1}

\begin{figure}%
    \centering
    \subfloat[\centering Model prediction: 0.01 (Normal)]{{\includegraphics[width=0.4\linewidth]{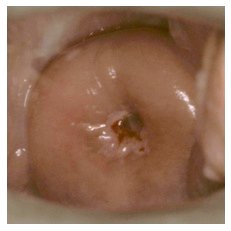} }}%
    \qquad
    \subfloat[\centering Model prediction: 0.98 (Pre-cancer)]{{\includegraphics[width=0.4\linewidth]{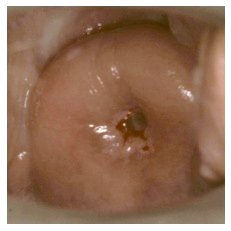} }}%
    \caption{\textbf{Illustration of repeatability issues from deep learning models on different images of a cervix with precancerous lesions from the same patient taken the same day.}  A binary model without dropout layers generated the following outputs. (a) the binary model predicts a normal cervix (severity score: 0.01). (b) the binary model predicts pre-cancer (severity score: 0.98).}%
    \label{fig:problem}%
\end{figure}

Deep learning is a popular technology to achieve high performance for medical image analysis tasks. In the desire to achieve higher classification performance, important aspects of the model performance such as test-retest variability remain overlooked,  yet not all deep learning (DL) models are equal with respect to their repeatability.
Consistency in the prediction of models is of utmost importance for such models to prove their potential as reliable and safe clinical support. However, DL models face substantial repeatability issues \citep{alahmari2020challenges, kim2020test}. Empirically, minor changes in an image can lead to vastly different predictions by DL models. In clinical practice, this repeatability issue could lead to dangerous medical errors. Figure \ref{fig:problem} illustrates this issue. Two cervical cancer screening images from the same precancerous cervix that were taken during the same visit lead to completely different predictions. A binary DL model (without dropout layers) trained to distinguish between a normal cervix and one with a precancerous lesion (0: Normal, 1: Pre-cancer) predicted a normal cervix on one image and classified the second image as precancerous. This difference is represented by prediction results at each extreme of the spectrum, i.e., 0.01 and 0.98, suggesting high certainty for both outputs.

Repeatability is an important and required characteristic of medial image analysis tools as it reflects the ability of the model to repeatedly generate a certain classification performance. More repeatable models with the same accuracy provide smaller variability in accuracy for a single measurement per patient. Hence, repeatable models generate more consistent classification performance leading to less variability.

\subsection{Study outline}
 While most works describing the development of DL models for medical image classification focus on accuracy and classification performance \citep{haenssle2018man, rajpurkar2018deep, bakas2018identifying}, very few assess the repeatability of these models. To the best of our knowledge, this is the first study to propose Monte Carlo (MC) dropout at test time as a method to improve repeatability and systematically assess this approach on different tasks, model types, and network architectures. All the selected medical tasks have an underlying continuous scale of disease severity but are routinely binned into binary or ordinal classes to simplify treatment decision and rating. Although specifically training networks to assess disease severity might be a preferred approach \citep{li2020siamese, heine2011quantitative, Campbell2016PlusVariability}, this is rarely done in practice \citep{thomas2020automated, lehman2019mammographic, brown2018automated}. The methodology and analysis were chosen based on the consideration that the underlying variable of interest, i.e. disease severity, of these medical tasks is better represented by a spectrum rather than clear distinct categories. In this work, we evaluate model repeatability of four types of DL models, binary classification, multi-class classification, ordinal classification, and regression, each with and without MC dropout. We test the repeatability of these models' predictions on four different medical image classification tasks: knee osteoarthritis grading, cervical cancer screening, breast density estimation, and retinopathy of prematurity (ROP) disease severity grading. True test-retest scenarios were studied with private datasets containing multiple images per patient for a given time point and anatomical region. Few public datasets exist with multiple images from the same anatomical region taken during the same visit. As we acknowledge the importance of reproducibility in research, a forth dataset that is publicly available, the Multicenter Osteoarthritis Study, was added to the study and a second image per patient was generated by applying simple data augmentation to the original image, i.e., horizontal flip, to simulate test-retest reliability. Based on our results, we present recommendations for model choices that can lead to improved repeatability. Finally, we assess the calibration of regular models compared to MC models.

\subsection{Related work}
\subsubsection{Dropout}
Dropout is the process of randomly removing units from a neural network during training to regularize learning and avoid overfitting \citep{hinton2012improving, srivastava2014dropout}. For inference, dropout is usually disabled to leverage all the connections from the model. Gal et al. \cite{gal2016dropout} proposed to enable dropout at test time as a Bayesian approximation to sample multiple different predictions. From these Monte Carlo (MC) predictions, it is possible to derive uncertainty metrics that are indicative of model performance \citep{camarasa2020quantitative} which has already been explored for multiple medical image classification tasks \citep{leibig2017leveraging, combalia2020uncertainty, singh2020skinet}. The final prediction is usually generated by taking the average over all MC predictions. We will refer to these models utilizing dropout as \textit{MC models}.

\subsubsection{Repeatability}
Repeatability describes the variation between independent tests taken under the same conditions. In this work, we focus on repeatability of a single model using different images of the same anatomical region from the same patient taken the same day. For the public knee osteoarthritis dataset, only one image per knee for a given time point was available, hence, a second image was generated using minor data augmentation. To the best of our knowledge, few studies focus on methodologies to increase repeatability. However, some work notes the importance of repeatability for medical image analysis by assessing the test-retest reliability of their classification or segmentation models \citep{kim2020test, hiremath2021test, estrada2020fatsegnet, cole2017predicting, hoebel2020radiomics, schwier2019repeatability, van2016repeatability}. Kim et al. \cite{kim2020test} evaluated the test-retest variability for disease classification on chest radiographs and obtained limits of agreement (LoA) of $~\pm 30\%$ indicating variability within the test re-test predictions. Various post-processing techniques such as blurring or sharpening, which could naturally occur in real-life settings and alter the appearance of images, caused higher test-retest variability compared to positional changes. Multiple other factors have been shown to impact repeatability such as inter-rater variability in the labels, image quality, noise, or model uncertainty due to lack of knowledge and limited number of images, i.e., epistemic uncertainty \citep{kim2020test, mojtahed2021repeatability}. For instance, images leading to high inter-rater variability among experts are likely to generate similar variability, especially at class boundaries \citep{Kalpathy-Cramer2016PlusAnalysis}, since the model was trained based on the ratings of these experts. While some of these factors leading to low repeatability cannot be eliminated in practice (e.g., inter-rater variability), reliable DL models should be robust to minor changes in position, lighting, focus, etc.

\subsubsection{Calibration}
Calibrated models will output probabilities reflecting the probability of the observed outcome (e.g., all the predictions of 0.9 from a perfectly calibrated model should have the positive class as ground truth 90\% of the time). Good calibration allows robust rejection of low probability predictions as output probabilities represent more truthfully the likelihood of being wrong. Modern neural networks are poorly calibrated due to the recent neural network advances in architecture and training \citep{guo2017calibration}. Multiple works have focused on developing methods for post-hoc calibration of models \citep{guo2017calibration, kuleshov2018accurate, laves2020well} usually taking the validation set to adjust the test prediction. However, having an inherently more calibrated output could mitigate the need of prediction re-calibration. Brier score is a common metric to assess calibration as it indicates how close the predicted probabilities are to the true likelihood. Brier score of 0 indicates perfect calibration. 

\section{Results}\label{sec2}

\subsection{Repeatability and classification performance}

\begin{figure}[htp]
  \centering
     \subfloat[\centering Knee osteoarthritis classification]{\includegraphics[height=0.2\textheight]{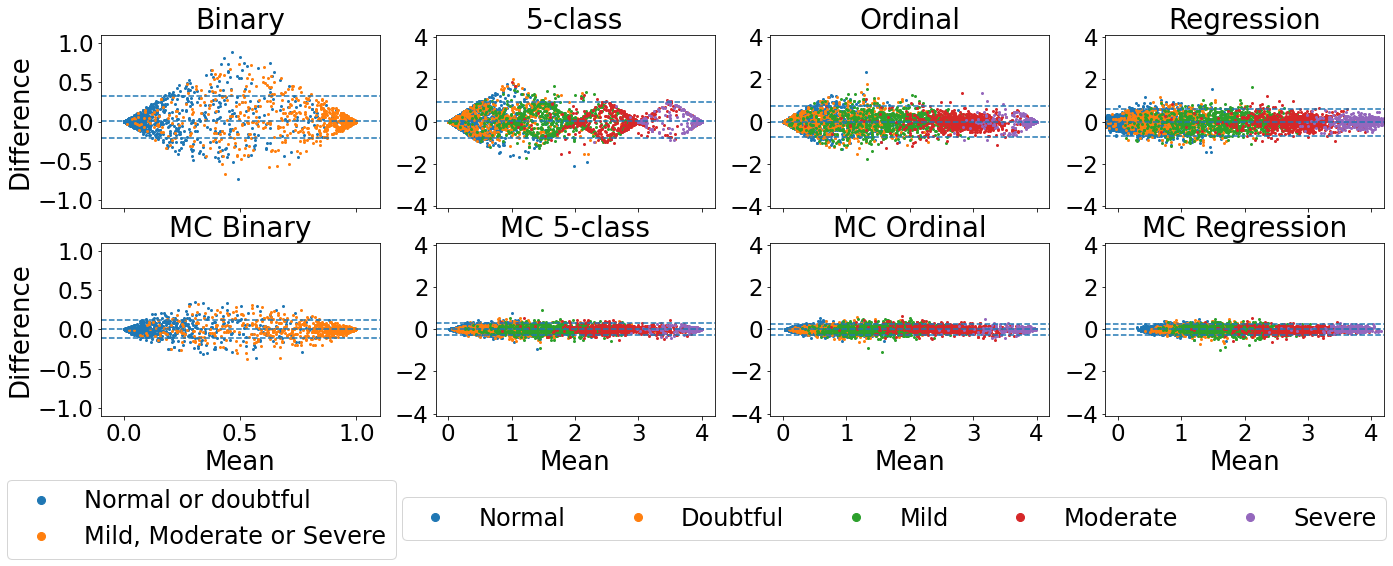} }
    \qquad
    \subfloat[\centering Cervical classification]{\includegraphics[height=0.2\textheight]{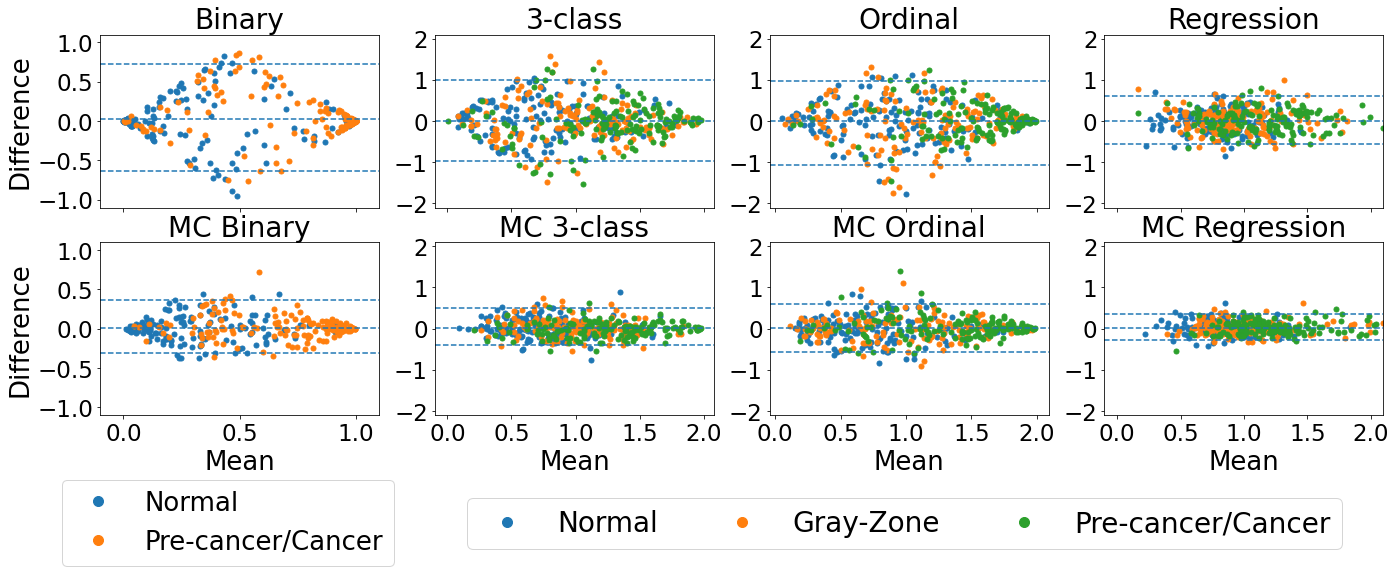} }
    \qquad
    \subfloat[Breast density classification]{\includegraphics[height=0.2\textheight]{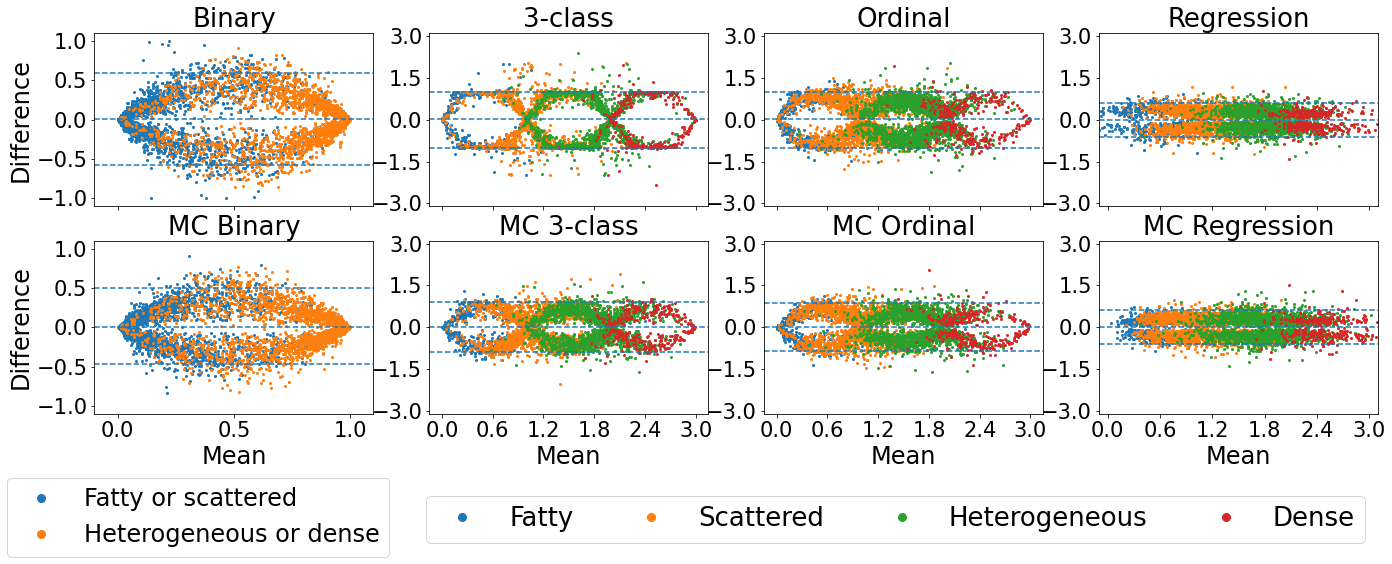}}
    \qquad
    \subfloat[ROP classification]{\includegraphics[height=0.2\textheight]{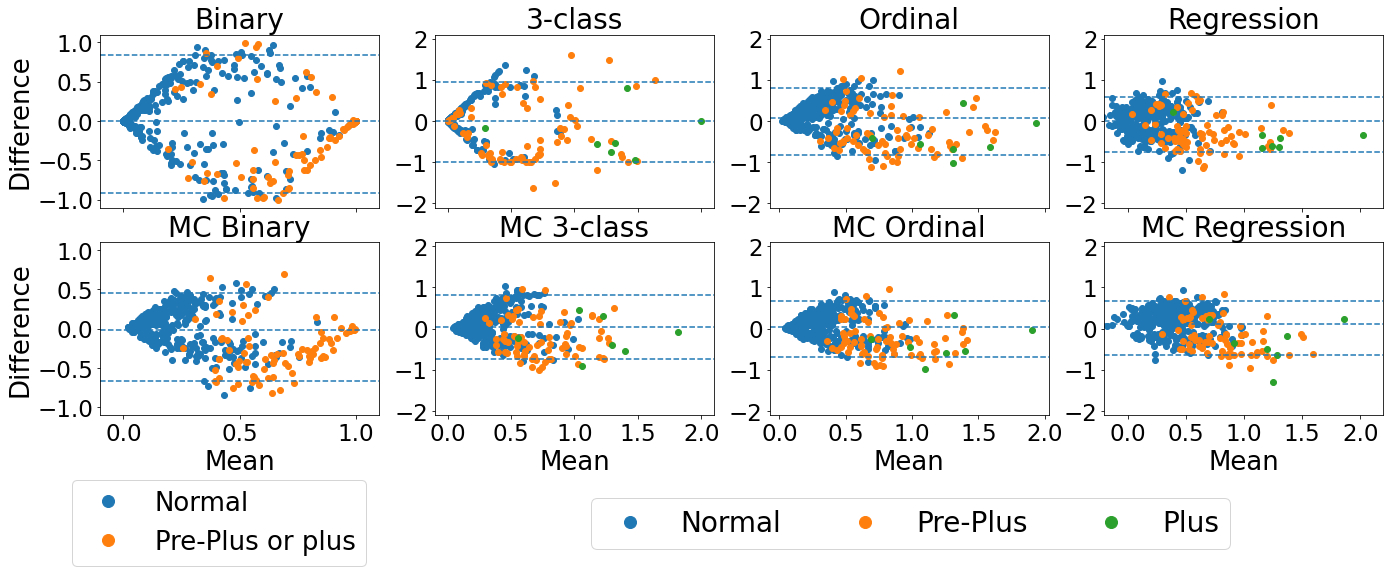}}

  \caption{\small{\textbf{Bland-Altman plots on multiple images from the same patient and visit.} The y-axis of each graph represents the maximum difference in model prediction for images of the same patient, while the x-axis refers to the mean of the predicted scores. The 95\% limits of agreement are presented with dashed blue lines. Repeatable models are associated with differences and limits of agreement closer to zero which indicates smaller difference between test and retest.}}
  \label{fig:bland_altman}
\end{figure}

The repeatability of each model was assessed on all available images of the same patient during the same visit. MC dropout models were associated with increased repeatability and accuracy for all models and tasks excluding regression models (Table \ref{tab:metric-table} and Figure \ref{fig:bland_altman}). Bland-Altman plots for all the tasks and model types are summarized in Figure \ref{fig:bland_altman}. An alternative way to compare the severity score from the test and retest images is presented in Appendix \ref{secA1}. Ideally, all cases would lie near a horizontal line crossing the y-axis at 0 which means the difference between test-retest score is low. For every task, the MC models showed better test-retest reliability than their conventional counterparts with the exception of the regression models. This is illustrated by the narrower 95\% LoA and the highest concentration of differences near 0 on the y-axis. Model outputs exhibit higher differences near class boundaries. However, this effect is attenuated for MC models and almost absent for regression models. The range of predicted values remained similar for MC models, indicating that the effect of the MC model is not simply regressing scores towards the mean. Moreover, the increase in repeatability was in most cases associated to an improvement in classification performance (Table \ref{tab:metric-table}).

Repeatability and classification metrics for each approach can be found in Table \ref{tab:metric-table}. Repeatability of MC models for binary, multi-class, and ordinal models showed statistically significant improvements on at least one metric for all tasks. On average, across all tasks and classification models (i.e., excluding regression), the disagreement rate improved by 7\% points and the 95\% limit of agreement by 16\% points. Classification performance followed the same trend as the repeatability and increased for all classification MC models with the exception of the ROP task which was exposed to a domain shift (see \ref{section:shift}). Adding MC iterations to regression models did not lead to consistent improvement in classification or repeatability performances. Regression models generally showed better repeatability compared to the other multi-class models (i.e., n-class and ordinal).

\begin{table}[hbtp]

    \centering
  
  {\caption{\small{\textbf{Model performance overview (MEAN $\pm$ 95\% CI)}. Values in bold indicate the best model between MC and non-MC models where a statistical difference ($p-value > 0.05$) was observed. The two first columns measure the model repeatability where smaller values indicate better repeatability. The two last columns represent the model performance and high values indicate better classification. Binary models were trained with the following classes: Knee osteoarthritis: none and doubtful vs. mild, moderate, and severe -- Cervix: normal vs. pre-cancer/cancer -- Breast density: fatty and scattered vs. heterogeneous and dense -- ROP: normal vs. pre-plus and plus. LoA: Limits of agreement; $\kappa$: Quadratic weighted Cohen's $\kappa$; Acc.: Accuracy; CI: Confidence interval.}}
  \label{tab:metric-table}
  }
  
  {\begin{tabular}{lcc|cc}
  \toprule
  & \multicolumn{2}{c}{\textbf{Repeatability metrics}} & \multicolumn{2}{c}{\textbf{Classification metrics}} \\
  \midrule
  \textbf{Model} & \textbf{\thead{Disag. \\ rate} $\downarrow$} & \textbf{95\% LoA} $\downarrow$  & \bm{$\kappa$} $\uparrow$ & \textbf{Acc.} $\uparrow$\\
   \midrule
  \multicolumn{5}{l}{\textbf{Knee osteoarthritis classification}} \\
  \midrule
  Binary & \small{$0.05 \pm  0.01$} & \small{$0.27 \pm 0.02$}  & \small{$0.87 \pm 0.01$} & \small{$0.95 \pm 0.00$} \\
  MC Bin. & \small{\bm{$0.02 \pm 0.00$}} & \small{\bm{$0.11 \pm 0.01$}} & \small{\bm{$0.89 \pm 0.01$}} & \small{\bm{$0.95 \pm 0.00$}} \\
 \arrayrulecolor{gray}\hline
  5-class & \small{$0.25 \pm 0.01$} & \small{$0.22 \pm 0.01$} & \small{$0.88 \pm 0.01$} & \small{$0.69 \pm 0.01$}\\
  MC 5-cl. & \small{\bm{$0.10 \pm 0.01$}} & \small{\bm{$0.07 \pm 0.00$}} & \small{\bm{$0.91 \pm 0.00$}} & \small{\bm{$0.72 \pm 0.01$}}\\
  \arrayrulecolor{gray}\hline
  Ord. & \small{$0.15 \pm 0.01$} & \small{$0.19 \pm 0.01$}  & \small{$0.84 \pm 0.01$} & \small{$0.54 \pm 0.01$} \\
  MC ord. & \small{\bm{$0.08 \pm 0.01$}} & \small{\bm{$0.07 \pm 0.00$}} & \small{\bm{$0.85 \pm 0.01$}} & \small{\bm{$0.56 \pm 0.01$}}\\
  \arrayrulecolor{gray}\hline
  Reg. & \small{$0.19 \pm 0.01$} & \small{$0.16 \pm 0.00$} & \small{\bm{$0.90 \pm 0.00$}} & \small{\bm{$0.70 \pm 0.01$}} \\
  MC Reg. & \small{\bm{$0.14 \pm 0.01$}} & \small{\bm{$0.07 \pm 0.00$}} & \small{$0.88 \pm 0.00$} & \small{$0.61 \pm 0.01$}\\
    \arrayrulecolor{black}
  \midrule
  \multicolumn{5}{l}{\textbf{Cervical classification}} \\
  \midrule
  Binary & \small{$0.23 \pm  0.05$} & \small{$0.68 \pm 0.07$} & \small{$0.46 \pm 0.07$} & \small{$0.73 \pm 0.03$} \\
  MC Bin. & \small{\bm{$0.13 \pm 0.04$}} & \small{\bm{$0.33 \pm 0.04$}} & \small{\bm{$0.51 \pm 0.07$}} & \small{\bm{$0.75 \pm 0.03$}}\\
 \arrayrulecolor{gray}\hline
  3-class & \small{$0.38 \pm 0.05$} & \small{$0.50 \pm 0.06$} & \small{$0.34 \pm 0.06$} & \small{$0.47 \pm 0.03$}\\
  MC 3-cl. & \small{\bm{$0.24 \pm 0.04$}} & \small{\bm{$0.22 \pm 0.03$}} & \small{\bm{$0.42 \pm 0.06$}} & \small{\bm{$0.52 \pm 0.03$}}\\
  \arrayrulecolor{gray}\hline
  Ord. & \small{$0.37 \pm 0.05$} & \small{$0.51 \pm 0.07$} & \small{$0.38 \pm 0.06$} & \small{$0.47 \pm 0.03$}\\
  MC ord. & \small{\bm{$0.28 \pm 0.04$}} & \small{\bm{$0.29 \pm 0.03$}} & \small{\bm{$0.41 \pm 0.06$}} & \small{\bm{$0.49 \pm 0.03$}}\\
  \arrayrulecolor{gray}\hline
  Reg. & \small{$0.31 \pm 0.04$} & \small{$0.29 \pm 0.03$} & \small{$0.34 \pm 0.05$} & \small{\bm{$0.44 \pm 0.03$}} \\
  MC Reg. & \small{\bm{$0.19 \pm 0.04$}} & \small{\bm{$0.16 \pm 0.02$}} & \small{\bm{$0.35 \pm 0.05$}} & \small{$0.43 \pm 0.03$}\\
    \arrayrulecolor{black}
  \midrule
  \multicolumn{5}{l}{\textbf{Breast density classification}} \\
  \midrule
  Binary & \small{$0.22 \pm 0.01$} & \small{$0.58 \pm 0.01$} & \small{$0.68 \pm 0.01$} & \small{$0.84 \pm 0.00$}\\
  MC Bin. & \small{\bm{$0.19 \pm 0.01$}} & \small{\bm{$0.48 \pm 0.01$}} & \small{\bm{$0.69 \pm 0.01$}} & \small{\bm{$0.85 \pm 0.00$}}\\
  \arrayrulecolor{gray}\hline
  4-class & \small{$0.54 \pm 0.02$} & \small{$0.33 \pm 0.00$} & \small{$0.71 \pm 0.01$} & \small{$0.69 \pm 0.01$} \\
  MC 4-cl. & \small{\bm{$0.45 \pm 0.01$}} & \small{\bm{$0.30 \pm 0.00$}} & \small{\bm{$0.72 \pm 0.01$}} & \small{\bm{$0.71 \pm 0.01$}} \\
  \arrayrulecolor{gray}\hline
  Ord. & \small{$0.52 \pm 0.01$} & \small{$0.33 \pm 0.00$} & \small{$0.70 \pm 0.01$} & \small{$0.68 \pm 0.01$} \\
  MC ord. & \small{\bm{$0.44 \pm 0.01$}} & \small{\bm{$0.29 \pm 0.01$}} & \small{\bm{$0.72 \pm 0.01$}} & \small{\bm{$0.69 \pm 0.01$}} \\
  \arrayrulecolor{gray}\hline
  Reg. & \small{\bm{$0.39 \pm 0.01$}} & \small{$0.21 \pm 0.01$} & \small{$0.74 \pm 0.01$} & \small{\bm{$0.70 \pm 0.01$}}\\
  MC Reg. & \small{$0.40 \pm 0.01$} & \small{$0.21 \pm 0.01$} & \small{\bm{$0.75 \pm 0.01$}} & \small{$0.67 \pm 0.01$}\\
  \arrayrulecolor{black}
  \midrule
  \multicolumn{5}{l}{\textbf{ROP classification}} \\
  \midrule
  Binary & \small{$0.31 \pm 0.01$} & \small{$0.88 \pm 0.04$} & \small{$0.50 \pm 0.05$} & \small{$0.81 \pm 0.02$} \\
  MC Bin. & \small{\bm{$0.25 \pm 0.04$}} & \small{\bm{$0.55 \pm 0.05$}} & \small{\bm{$0.56 \pm 0.05$}} & \small{\bm{$0.85 \pm 0.02$}}\\
  \arrayrulecolor{gray}\hline
  3-class & \small{$0.23 \pm 0.04$} & \small{$0.48 \pm 0.03$} & \small{\bm{$0.57 \pm 0.06$}} & \small{$0.85 \pm 0.02$}\\
  MC 3-cl. & \small{$0.23 \pm 0.04$} & \small{\bm{$0.39 \pm 0.03$}} & \small{$0.55 \pm 0.06$} &  \small{\bm{$0.85 \pm 0.02$}}\\
  \arrayrulecolor{gray}\hline
  Ord. & \small{$0.31 \pm 0.04$} & \small{$0.40 \pm 0.04$} & \small{$0.57 \pm 0.05$} & \small{$0.82 \pm 0.02$}\\
  MC ord. & \small{\bm{$0.29 \pm 0.04$}} & \small{\bm{$0.34 \pm 0.03$}} & \small{$0.57 \pm 0.05$} & \small{\bm{$0.83 \pm 0.02$}} \\
  \arrayrulecolor{gray}\hline
  Reg. & \small{\bm{$0.16 \pm 0.04$}} & \small{\bm{$0.33 \pm 0.03$}} & \small{\bm{$0.58 \pm 0.06$}} & \small{\bm{$0.86 \pm 0.02$}}\\
  MC Reg. & \small{$0.47 \pm 0.05$} & \small{$0.33 \pm 0.01$} & \small{$0.51 \pm 0.05$} & \small{$0.79 \pm 0.03$}\\
  \arrayrulecolor{black}
  \bottomrule
  \end{tabular}}
\end{table}

\subsection{Impact of number of MC iterations}
\begin{figure}[htp]
  \centering
      \subfloat[\centering Knee osteoarthritis 5-class model]{\includegraphics[width=0.39\textwidth]{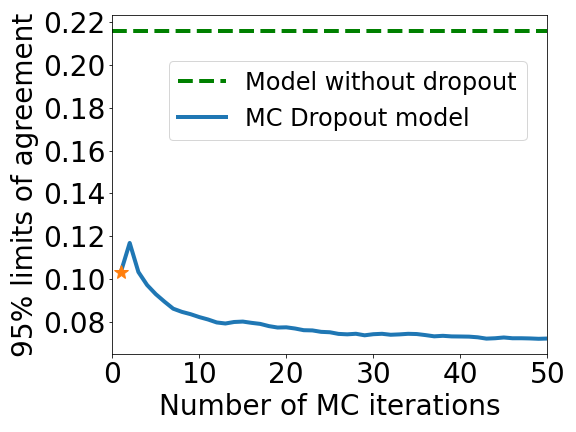} }
    \qquad
    \subfloat[\centering Cervical 3-class model]{\includegraphics[width=0.39\textwidth]{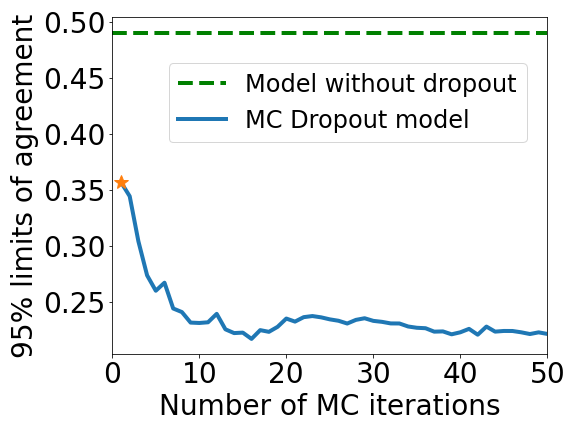} }
    \qquad
    \subfloat[Breast density 4-class model]{\includegraphics[width=0.39\textwidth]{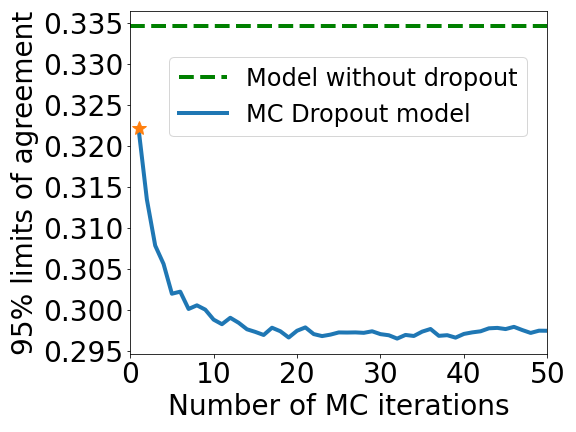}}
     \qquad
        \subfloat[\centering ROP 3-class model]{\includegraphics[width=0.39\textwidth]{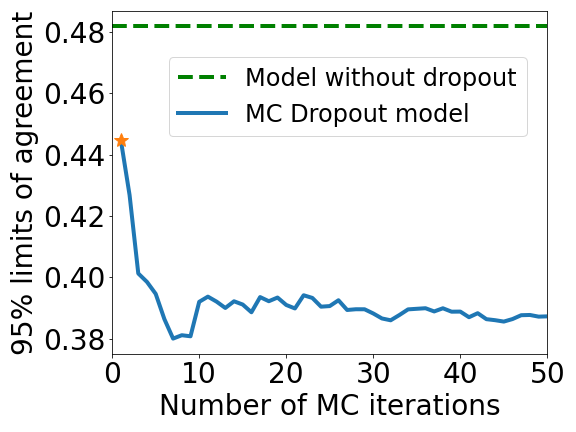} }

  \caption{\textbf{Impact of number of MC iterations on repeatability.} The orange star represents a single forward pass of the model with dropout disabled at test time. }
  \label{fig:mc_it}
\end{figure}

Additionally, we evaluated the impact of increasing the number of MC iterations at test time to compute the final prediction on repeatability of MC models, i.e., 95\% LoA, of multi-class models for all tasks as illustrated in Figure \ref{fig:mc_it}. This analysis was limited to the multi-class models as they are the most commonly used for medical classification tasks. All models suggest that training with dropout, even without any MC iterations during testing, has better test-retest performance than non-dropout models (Figure \ref{fig:mc_it}). Repeatability could be further improved by generating more MC samples. After about 20 MC iterations, additional samples had little to no impact on repeatability.

\subsection{Architecture comparison}
\begin{figure}[htp]
  \centering
    \includegraphics[height=0.2\textheight]{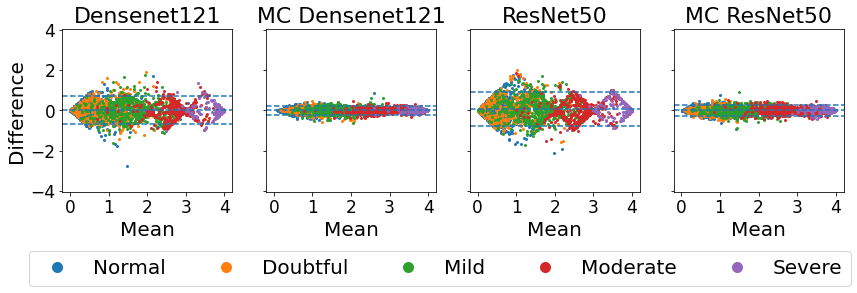}

  \caption{\textbf{Architecture comparison on multi-class model for knee osteoarthritis grading.} The two first columns are the model trained with Densenet121 while the two last ones were trained with ResNet50. The first and third graphs represent the regular model and the second and fourth ones display their MC counterparts. }
  \label{fig:architecture}
\end{figure}

Figure \ref{fig:architecture} compares, for the same task (i.e., knee osteoarthritis grading) and model type (i.e., multi-class), the DenseNet and ResNet architectures with respect to repeatability. Regardless of the model's architecture, the behavior remains the same: the test-retest variability is lower meaning repeatability is increase when using multiple MC samples for the prediction. The disagreement rate decreased of 9\% and 15\% points and the LoA improved by 11\% and 15\% points for DenseNet and ResNet architectures respectively.

\begin{figure}[htp]
  \centering
    \subfloat[\centering Knee osteoarthritis 5-class models]{\includegraphics[width=0.46\textwidth]{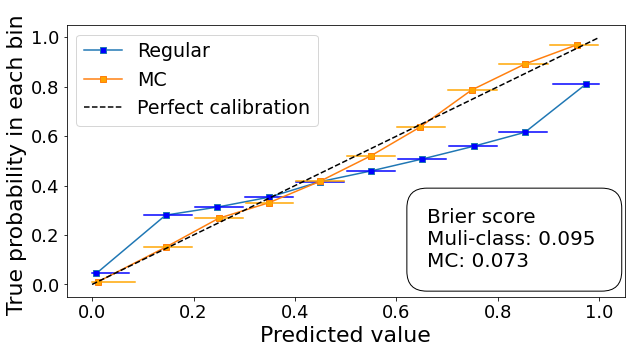} }
    \qquad
    \subfloat[\centering Cervical 3-class models]{\includegraphics[width=0.46\textwidth]{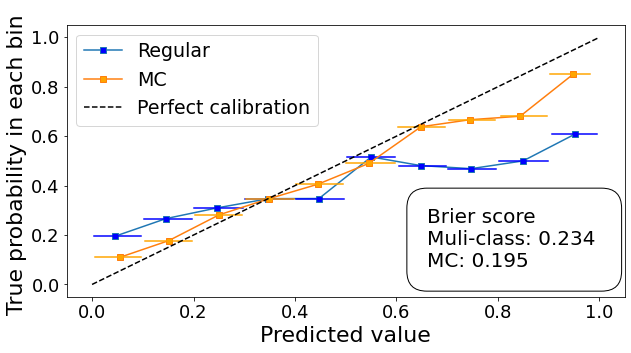} }
    \qquad
    \subfloat[Breast density 4-class models]{\includegraphics[width=0.46\textwidth]{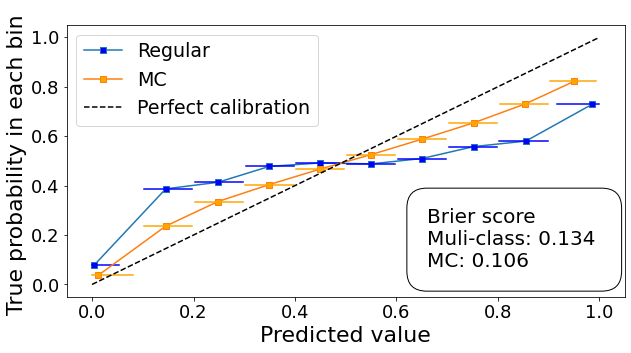}}
    \qquad
    \subfloat[ROP binary models]{\includegraphics[width=0.46\textwidth]{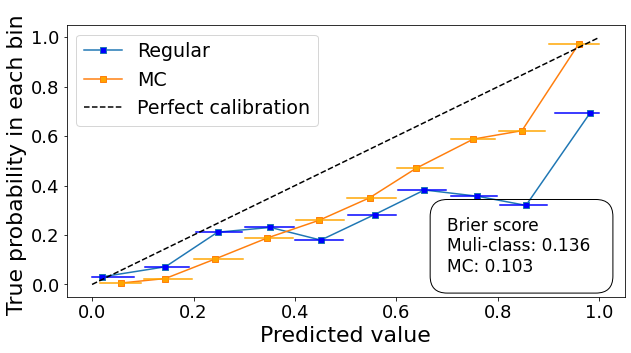}}

  \caption{\textbf{Calibration curves.} Brier score quantifies model calibration: 0 indicates a perfectly calibrated model. The horizontal bars represent the predicted value distribution (95\% CI) for every bin.}
  \label{fig:calibration}
\end{figure}

\subsection{Calibration}
Output probabilities are more calibrated for MC models than for the regular models as depicted in Figure \ref{fig:calibration}. Brier scores associated with MC models are lower for all tasks, i.e., average decrease of 0.031, and the calibration curves are closer to the identity line, i.e., the perfect calibration curve. Calibration curves of multi-class model outputs were displayed for knee osteoarthritis, cervix and breast density classification while the binary models were chosen for ROP as the impact of adding MC was greater for this task compared with the multi-class models (see Table \ref{tab:metric-table}).

\section{Discussion}\label{sec3}

\subsection{Repeatability and accuracy} Our results demonstrate that MC dropout models lead to a significant increase in repeatability, i.e., improvement of at least one repeatability metric, while improving most classification metrics for binary, multi-class, and ordinal models. Concretely, this means higher class and score agreements between the test and retest outputs. The repeatability increased regardless of the disease imaged or the model architecture (DenseNet or ResNet). However, MC iterations did not benefit regression models and even lowered classification performance for knee osteoarthritis and ROP classification. Regression models showed higher repeatability compared with non-MC multi-class and ordinal models, so the potential gain was more modest. While the lowest test-retest variability was reached for the regression model on the knee and cervical images, the model was associated with a lower quadratic $\kappa$ and/or accuracy. Both accuracy and repeatability need to be reported to thoroughly assess deep learning models, especially in clinical settings. 

\subsection{Bland-Altman plots and calibration} The observed differences between test-retest images of the same patient was not constant along the mean axis as seen on the Bland-Altman plots in Figure \ref{fig:bland_altman}. Near the class boundaries, images show more variability with only a few cases with a difference near zero, which creates an arch-like pattern in the plots. This phenomenon can be partly explained by the training scheme of classification models. During training, models are optimized to predict classes with high certainty, discouraging the model to output ambivalent predictions (e.g., predicting 0.5 for a binary model) which leads to uncalibrated models \citep{guo2017calibration}. Ideally, the output softmax or sigmoid probability of a model should reflect the uncertainty of the model between two or more classes. However, in practice, this is not the case leading to high differences on the class boundaries due to misclassification of at least one of the images. This effect is alleviated with MC models, leading to more calibrated outputs and higher repeatability.

% \paragraph{} Sampling multiple predictions with dropout generate a score estimation 

\subsection{Domain shift}
\label{section:shift} 
Fewer repeatability metrics showed a statistical difference between MC dropout and conventional models for the ROP disease severity classification task. Unlike knee osteoarthritis, cervical and breast density classification, the ROP models were tested on views of the eye that the model has not seen during training (section \ref{section:rop}). This domain shift might be adding variability in the model's prediction impacting the global performance and repeatability, effectively abating the benefits of MC dropout models. Nonetheless, MC models still showed a higher repeatability under domain shift than no-dropout models.

\subsection{Number of MC iterations}  MC models are computationally more expensive than their conventional counterparts as they require multiple forward passes at testing time. Our results on Figure \ref{fig:mc_it} indicate that after approximately 20 MC iterations, there is no further gain in repeatability, and this, for all tasks on multi-class models. For settings were time and computational resources are limited, training with dropout layers, even without sampling multiple MC, helps regularize the training and reduces overfitting \citep{hinton2012improving}.

\section{Methods}\label{sec4}

\subsection{Datasets}

All images were de-identified prior to data access, ethical approval for this study was therefore not required.

\subsubsection{Knee osteoarthritis}
\paragraph{Knee osteoarthritis - background}
Knee osteoarthritis is the most common musculoskeletal disorder \citep{tiulpin2018automatic} and was the eleventh highest contributor to global disability in 2010 \citep{cross2014global}. Osteoarthritis can be diagnosed with a radiography, however, early diagnosis can be challenging in clinical practice and is prone to inter-rater variability justifying the emergence of AI models for osteoarthritis grading \citep{tiulpin2018automatic}. The severity is typically measured using the Kellgren-Lawrence (KL) scale from 0 to 4 where 0 corresponds to none, 1 to doubtful, 2 to mild, 3 to moderate and 4 to severe \citep{kellgren1957radiological}. 

\paragraph{Knee osteoarthritis - dataset description}
The publicly available longitudinal Multicenter Osteoarthritis Study (MOST) dataset contains 18 926 knee radiographies from 3017 patients of one or both knees when including only grades from 0 to 4 on the Kellgren-Lawrence scale \citep{kellgren1957radiological}. Grades outside the Kellgren-Lawrence scale were excluded from the dataset for this work.  40\% of the cases were labelled as grade 0, 15\% as grade 1, 17\% as grade 2, 19\% as grade 3 and 9\% as grade 4. The patients were split into training, validation, and test sets representing 65\%, 10\%, and 25\% of the images respectively. The binary models were trained to distinguish between knees with no or doubtful osteoarthritis (negative class) and knees with mild, moderate or severe osteoarthritis (positive class). Images were center cropped to a size of 224x224 pixel and scaled to intensity values of 0 to 1. MOST does not include multiple images of the same during the same visit. Model predictions were generated for all the original test images, were then flipped horizontally, and re-tested to emulate a test-retest setting. Hence, the repeatability was measured on the same radiography from the same patient at a given time point with and without the horizontal flip. 

% 3017 patients,  18926 images
% 0.0    7649
% 3.0    3538
% 2.0    3187
% 1.0    2851
% 4.0    1701

% 0.0    0.404153
% 3.0    0.186939
% 2.0    0.168393
% 1.0    0.150639
% 4.0    0.089876

% train    12268
% test      4753
% val       1905

% train    0.648209
% test     0.251136
% val      0.100655

\subsubsection{Cervical}
\paragraph{Cervical cancer screening - background}
Cervical cancer is the fourth most common cancer world wide and the leading cause of cancer-related deaths of women in western, eastern, middle, and southern Africa \citep{Arbyn2020EstimatesAnalysis}. 
Vaccinations against high risk strains of the Human Papilloma Virus (HPV) have been proven to prevent up to 90\% of cervical cancers \citep{Lei2020HPVCancer}. Until HPV vaccination programs have not reach every eligible woman worldwide and in light of the high the prevalence of high risk HPV types, there will be a great demand for effective screening at low costs to prevent the development of invasive cervical cancer. 
In addition to HPV testing, the visual assessment of the cervix using photographs can help to detect precancerous lesions in low-resource settings \citep{Catarino2015CervicalChoices, Xue2020ACamera, Hu2019AnScreening}.

\paragraph{Cervigram - dataset description}
The cervical cancer screening dataset consisted of 3509 cervical photographs from 1760 patients from two studies \citep{bratti2004description, schiffman2003findings}. For most patients, we had access to two cervical photographs taken during the same session.

Each image was classified using cytological and histological data from the patient as one of the following three categories: Normal (1148 images, 33\%),  Gray zone, i.e., the presence of precancerous lesions was equivocal, (1159 images, 33 \%), Pre-cancer/cancer (1202 images, 34\%).

The dataset was split into training (65\%), validation (10\%), and test sets (25\%) on a patient level, resulting in datasets containing 2283, 350, and 876 images (training/validation/test) preserving the class distributions described above within each subset.
All images were de-identified before this study. 
All cervical images were cropped using bounding boxes from a trained Retina net for cervix detection, resized to 256x256 pixel, and scaled to intensity values of 0 to 1.
The cervigram classification models were trained using all photographs for each patient in the training dataset. 
For the binary classification models, we utilized only images that were classified as either normal or pre-cancer/cancer.
For all patients in the test dataset for whom both images were available, repeatability was assessed as the difference in predictions between the two photographs. 

\subsubsection{Breast density}
\paragraph{Breast density classification - background} 
Breast cancer is the second most common cause of cancer deaths among women in the USA with an estimated number of more than 41,000 deaths in 2019 \citep{Siegel2019Cancer2019}.
The density of a women's breast is determined by the amount of fibroglandular tissue. It can be classified (with increasing density) based on its appearance on x-ray mammography as almost entirely fatty, scattered fibroglandular densities, heterogeneously dense, and extremely dense \citep{Liberman2002BreastBI-RADS}. 
Importantly, the risk to develop breast cancer rises with increasing breast density \citep{Boyd1995QuantitativeStudy}. 
Furthermore, \cite{Bakker2019SupplementalTissue} have shown that women with extremely dense breast tissue benefit from additional MRI screening. The development of AI models based on expert labels for breast density assessment could help to mitigate intra-, and interobserver variability and the inconsistency of current quantitative measurements with expert raters \citep{lehman2019mammographic}.

\paragraph{DMIST - dataset description} 
The Digital Mammographic Imaging Sceening Trial (DMIST) dataset consists of a total of 108,230 mammograms from 21,729 patients acquired at 33 institutions with an average of five mammographs of different standard mammography views for each patient \citep{Pisano2005DiagnosticScreening}. 
Breast density labels were generated according to the BI-RADS criteria \citep{Liberman2002BreastBI-RADS} by a total of 92 different radiologists.
The dataset consisted of 12,428 (11.5\%) fatty, 47,909 (44.2\%) scattered, 41,325 (38.2\%) heterogeneously dense, and 6,568 (6.1\%) extremely dense samples and was split into training (70,293), validation (10,849), and test datasets (27,048  images) on a patient level preserving the label distribution of the full dataset.
All images were de-identified before this study. 
We cropped all images to a size of 224x224 pixels.
The breast density classification models were trained using all available views for each patient in the training dataset using either four labels or a simplified binary labelling system of fatty and scattered as one class, and dense and heterogeneous as the other class.
Repeatability was assessed as the maximum difference between all available views for each patient in the test dataset. 

\subsubsection{Retinopathy of Prematurity}
\label{section:rop}
\paragraph{Retinopathy of prematurity - background}
ROP is the leading cause of preventable childhood blindness worldwide \citep{BlindnessVISION2020}. It gets diagnosed based on the appearance of the retinal vessel tree on retinal photographs and classified into three discrete disease severity classes: normal, pre-plus, and plus disease \citep{Quinn2005ThePrematurity}. However, the disease spectrum is continuous \citep{Campbell2016PlusVariability} and the use of discrete class labels to train DL classifiers is complicated by inter-rater variability particularly for cases close to the class boundaries \citep{Kalpathy-Cramer2016PlusAnalysis, Chiang2007InterexpertPrematurity}. High interrater variability, an insufficient number of ophthalmologists and neonatologists with the expertise and willingness (e.g., due to significant malpractice liability) to manage ROP, and the rising incidence of ROP worldwide motivate the development of AI models for ROP classification and screening \citep{Brown2018AutomatedNetworks}.

\paragraph{ROP - dataset description}
The ROP dataset consists of 5511 retinal photographs acquired at eight different study centers \cite{Brown2018AutomatedNetworks}. 
For each patient, retinal photographs were acquired in 5 different standard fields of view (posterior, nasal, temporal, inferior, superior). 
Only the posterior, temporal, and nasal views were used in this study. 
Images were classified as normal, pre-plus disease, or plus disease following previously published methods \citep{Ryan2014DevelopmentOphthalmology}. The final label is based on the independent image-based diagnosis by 3 expert graders in combination with the full clinical diagnosis by an expert ophthalmologist. Of the 5511 images in the dataset 4535 (82.3\%) were classified as normal, 804 (14.6\%) as pre-plus disease, and 172 (3.1\%) as plus disease. The binary models were trained to distinguish between normal and pre-plus/plus disease.
The dataset was split on a patient level into training, validation, and test datasets containing 4322/722/467 images while preserving the overall class distribution within each subset. Following \cite{Brown2018AutomatedNetworks}'s work, we trained ROP classification models using normalized pre-segmented vessel maps as input (size of 480x640). 
ROP classification models were trained using only the posterior field of view as ROP refers to arterial tortuosity and venous dilation within the posterior pole of the retina \citep{campbell2016expert}. However, it was shown that experts use characteristics beyond the posterior view to assess ROP severity\citep{campbell2016expert}. Hence, repeatability was tested using the posterior, temporal, and nasal views of all patients in the test dataset.

\subsection{Classification model training}
For each dataset, we trained binary, multi-class and ordinal \citep{cao2019rank} classification models, as well as regression models each with and without dropout, resulting in a total of 8 models per dataset. 
Models with dropout were trained using spatial dropout with a dropout rate of 0.1 for cervial images and DMIST, and 0.2 for knee osteoarthritis and ROP. The dropout rates were determined based on preliminary explorations to optimize the model's classification performance. For the DenseNet121 architecture, the dropout was applied after every dense layer while for the ResNets the dropout layer was applied after each residual block. At test time, the dropout was enabled to generate $N=50$ slightly different predictions and the final prediction was obtained by averaging over all the MC samples \citep{gal2016dropout}. The choice of the number of MC predictions was based on values commonly found in the appropriate literature and experience; however, the optimal number of predictions to reach maximum repeatability was assessed in the results section (see Figure \ref{fig:mc_it}).
We used the following ImageNet pretrained models for each dataset based on which performed the best for the conventional multi-class classification model: DenseNet121 (cervix), ResNet50 (knee osteoarthritis, breast density), and ResNet18 (ROP). Models were trained using binary cross-entropy, cross-entropy, CORAL \citep{cao2019rank}, and mean squared error (MSE) losses for binary, multi-class, ordinal, and regression models respectively. Affine transformations i.e., rotation $\pm 15$ degrees and random flips with 50\% probability, were applied as data augmentation during training. The code was implemented using the MONAI framework \citep{the_monai_consortium_2020_4323059} based on the PyTorch library \citep{NEURIPS2019_9015}.

\subsection{Evaluation}
\subsubsection{Severity scores} 

For direct comparison of a model's predictions, we summarized each model's outputs as a continuous severity score. For the binary and regression models, the output of the models was directly used without further modifications. For the multi-class model, we utilize the ordinality of all four classification problems and defined the continuous severity score as a weighted average using softmax probability of each class as described in Equation \ref{eq:multiclass}. For knee osteoarthritis (5 classes), the values lie in the range of 0 to 4, for breast density (4 classes) in the range of 0 to 3, and for cervical and ROP classification (3 classes), in the range 0 to 2.

 \begin{equation} \label{eq:multiclass}
    score = \sum_{i=1}^{k} p_{i} \times i - 1
 \end{equation}

\noindent with $k$ being the number of classes and $p_{i}$ the softmax probability of class $i$.
For the ordinal model, the classification problem of $k$ ranks (i.e., class) is modified into a $k - 1$ binary classification \citep{li2007ordinal} leading to one output unit less than for the traditional classification model. For instance, for a 3-class problem, the ground truth would be encoded as followed: class 1 $\rightarrow$ [0, 0]; class 2 $\rightarrow$ [1, 0]; class 3 $\rightarrow$ [1, 1]. The continuous prediction score for ordinal models is obtained by summing the output neurons. Similarly to the multi-class models, values range from 0 to 2, 0 to 3, and 0 to 4, for 3-class, 4-class, and 5-class problems respectively.

\subsubsection{Metrics} 
Repeatability was evaluated using the classification disagreement rate and the 95\% LoA from the Bland-Altman plots. Since normality was not reached for the differences for the LoA, non-parametric LoA were calculated using empirical percentiles \citep{bland1999measuring}. The LoA was presented as a fraction of the possible value range. The classification disagreement rate corresponds to the proportion of patients with different classification outcomes for different images acquired during the same session over the total number of patients. The classification accuracy and quadratic weighted Cohen's $\kappa$ were also reported. For the regression models, thresholds to binarize predictions for accuracy and Cohen's $\kappa$ calculation were computed by splitting the range of predictions equally (e.g., 3-class problem: $s \leq 0.67 \rightarrow$ class 1; $0.67 < s \leq 1.33 \rightarrow$ class 2; $s \geq 1.33 \rightarrow$ class 3). Model calibration was assessed using Brier score.

\subsubsection{Statistical analysis}
Statistical difference between models was determined using a two-sided $t$-test and metric bootstrapping (500 iterations). Models with a p-value smaller than 0.05 were considered significantly different. The normality of the distribution was verified using the Shapiro-Wilk test ($\alpha=0.05$).

\section{Conclusion}\label{sec5}
We evaluated the repeatability of four model types on four medical tasks using distinct model architectures (ResNet18, ResNet50, DenseNet121). We demonstrated that MC sampling during test time leads to more reliable models providing more stable, repeatable and calibrated predictions on different images from the same patient with or without a slight domain shift. Only regression models did not show a constant improvement when leveraging MC sampling. Repeatability metrics increased with an increasing number of MC iterations; after around 20 MC iterations, no further improvement of repeatability could be reached. MC sampling is flexible as it is applicable to any model type and architecture while being easily implementable. Future work should assess the impact of MC models on repeatability for other model architectures and other tasks such as segmentation.

\section{Data availability}\label{sec6}
Access to the MOST dataset for knee osteoarthritis can be requested through the NIA Aging Research Biobank \url{https://agingresearchbiobank.nia.nih.gov/}. The cervical, breast density, and ROP datasets are not publicly accessible due to patient privacy restrictions.

\section{Code availability}\label{sec7}
The code used to train and generate results can be found at \url{https://github.com/andreanne-lemay/gray_zone_assessment}.

\section{Funding}
Funded by the National Institutes of Health (Bethesda, MD) [R01 HD107493], an investigator-initiated grant from Genentech (San Francisco, CA) [R21 EY031883], and by unrestricted departmental funding and a Career Development Award (JPC) from Research to Prevent Blindness (New York, NY) [P30 EY10572]. A.L. has a scholarship from Mitacs [IT24359], NSERC, and “Fondation et Alumni de Polytechnique Montréal”.  B.B. has a fellowship from NCI/NIH [T32CA09168].

\section{Competing Interests}
The authors declare no competing financial or non-financial interests.

\section{Author contributions}
Study concept and design: A.L., K.H., C.P.B., J.K.-C. Data collection: S.D.S, A.C.R., and M.S. for cervical data, and J.P.C and J.K.-C. for ROP data. Data analysis and interpretation: all authors. Drafting of the manuscript: A.L., K.H. Critical revision of the manuscript for important intellectual content and final approval: all authors. Supervision: J.K.-C., K.H.
\newpage
\begin{appendices}

\section{Test retest severity score visualization}\label{secA1}

An alternative way to present the data from Figure \ref{fig:bland_altman} is to directly plot the severity score from a test image compare to the prediction obtained during retest. Figure \ref{secA1} displays the relation between pairs of images from the same patient taken at a given time point for all model types for knee osteoarthritis and cervical classification. When more than two images were available, the pair associated with the largest difference was selected. The data points are expected to lie near the identity line where the severity scores are the equal. As seen in Figure \ref{fig:bland_altman}, MC models exhibit increased repeatability compared with non-MC models as the data points are more concentrated near the identity line.

\begin{figure}[htp!]
  \centering
     \subfloat[\centering Knee osteoarthritis classification]{\includegraphics[height=0.2\textheight]{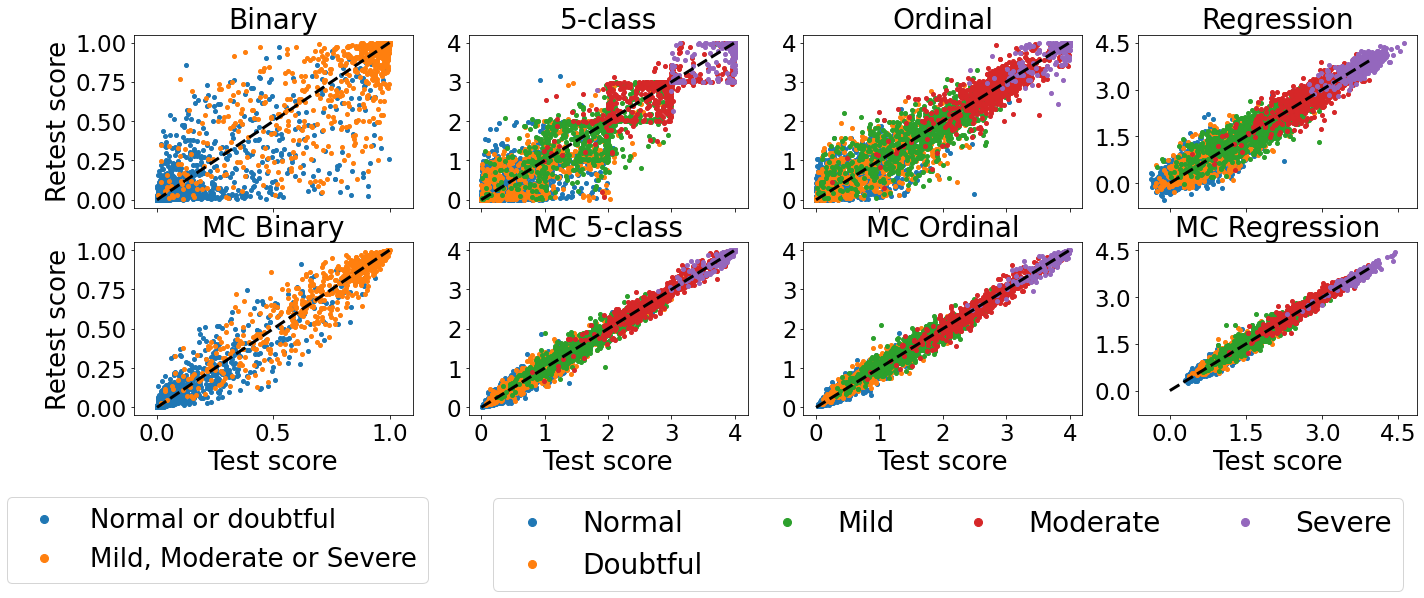} }
    \qquad
    \subfloat[\centering Cervical classification]{\includegraphics[height=0.2\textheight]{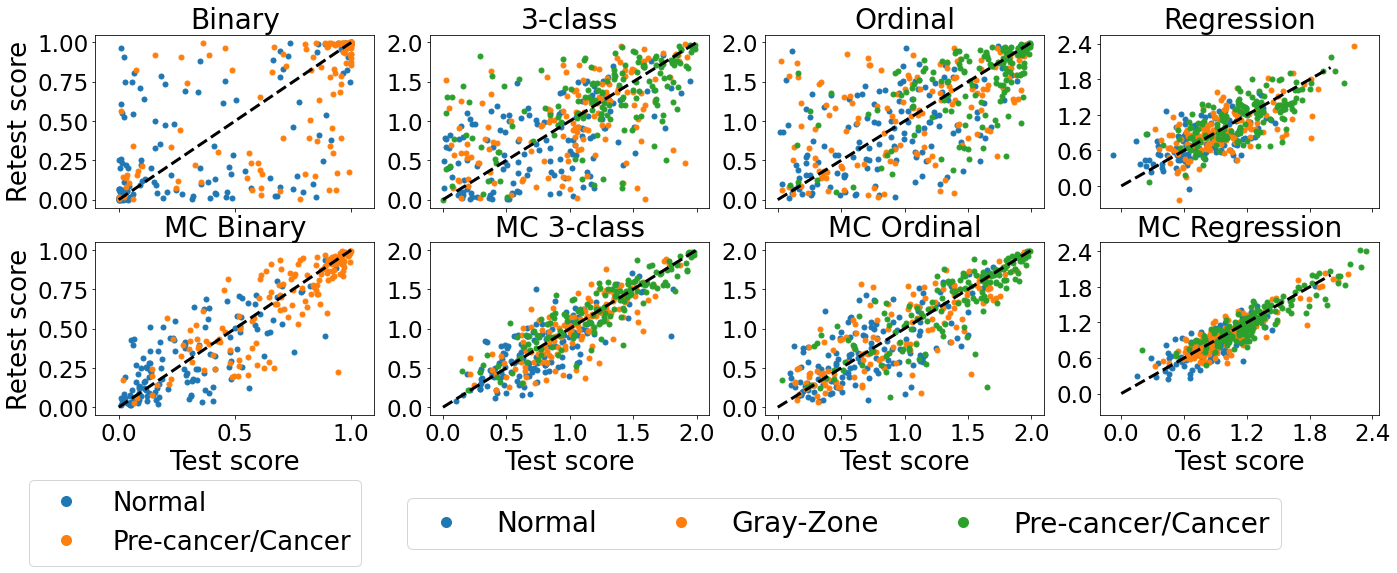} }

  \caption{\small{\textbf{Comparison of severity scores obtained on different images, test and retest, from the same patient taken during the same visit.} Each data point is a pair a severity score from test retest images. When more than two images were available, the pair with the largest difference was retained. The dash line represents the identity line where the image pair are expected to be since the severity score from both images should have the same value.}}
  \label{fig:img1vs2}
\end{figure}

\end{appendices}

\newpage
\bibliography{sn-bibliography}

\end{document}